\documentstyle[epsfig,aps]{revtex}

\newcommand{\be}{\begin{eqnarray}}
\newcommand{\ee}{\end{eqnarray}}

\newcommand{\bea}{\begin{eqnarray}}
\newcommand{\eea}{\end{eqnarray}}

\begin{document}

\thispagestyle{empty}

\title{Ion induced  quark-gluon implosion}
\author{L.~Frankfurt$^a$,  M.~Strikman$^b$\\
 {\small\it $^a$School of Physics and Astronomy, 
 Tel Aviv University, Tel Aviv, Israel}\\
 {\small\it $^b$Department of Physics, Pennsylvania State University,
                University Park, PA 16802, USA}\\
}

\maketitle
\begin{abstract}
We investigate nuclear fragmentation in the central proton-nucleus and
nucleus - nucleus collisions at the energies of CERN LHC. 
Within
the semiclassical approximation
we argue that
each nucleon is stripped in the average process off ``soft'' partons and 
fragments into a collection of leading quarks and gluons with large $p_t$
because of the fast increase with energy of the cross sections of soft 
and hard interactions. 
Valence quarks and gluons are streaming in the opposite 
directions when viewed in the  c.m.  of the produced system. The resulting 
pattern of the fragmentation of the colliding nuclei leads to an implosion  
of the quark and gluon constituents of the nuclei. The non-equilibrium state 
produced at the initial stage  in the nucleus fragmentation region 
is estimated to have densities  $\geq$ 50 GeV/fm$^3$ at the LHC energies
and probably $\geq$ 10 GeV/fm$^3$ at BNL RHIC.                                
\end{abstract}
One of challenging theoretical phenomena is the fact that 
perturbative QCD interactions become strong at sufficiently small x
due to fast increase with energy of gluon densities. 
A number of models were suggested for 
the theoretical description of this new QCD regime 
see, e.g. \cite{McDermott,Muellerrev,Iancu:2002xk} for reviews. A 
common feature
of all these models is that in the case of proton scattering off heavy
nuclei with  large radius $R_A$, the high-energy cross section
for most important quark-gluon configurations in the projectile nucleon    
reaches the black body limit(BBL) $\sigma_{tot}=2\pi R_A^2$ 
\cite{footnote1}. They have this feature
because, for the majority of the quark-gluon components of the nucleon
wave function, the interaction cross section satisfies the condition: 
$\sigma 2\rho R_A\gg 1$. Therefore, most components are absorbed by
the nuclear target. Here $\rho \approx$ 0.16 fm$^{-3}$ is the nuclear density. 
For the central nucleus-nucleus  collisions, the 
cross section of hard interaction is even more enhanced by the 
factor $\left[g_A(x,Q^2)/ \pi R_A^2\right]/ (g_N/\pi r_N^2)
\approx A^{1/3}(g_A/Ag_N)\approx 6$ for $A\approx 200$. Part of the
enhancement disappears because of the leading twist nuclear shadowing 
phenomenon. Evidently, the soft QCD interactions in the
nucleon-heavy nucleus collisions at the CERN Large Hadron Collider (LHC) 
should  also be close to the BBL because of
$\sigma_{tot}(pN)\approx 100 mb$.

The numerical estimates indicate that already for $x\sim 10^{-4}$ the 
perturbative cross section of the interaction
of $q\bar q $ dipoles of transverse sizes $d \geq $0.15fm with nuclei 
calculated within the framework of the QCD factorization theorem 
reaches  BBL while for the octet dipoles of similar sizes this  regime 
may start at x about 10 times larger.  For the recent numerical studies 
see \cite{FGSrev}. Hence,  the proton (ion) - heavy ion collisions at LHC will
be qualitatively different from those at fixed target energies and  
as well as at BNL Relativistic Heavy Ion Collider (RHIC) where  typical 
$x\geq 10^{-2}$ for  the central rapidities. This should lead to a number of 
novel soft and hard phenomena \cite{FELIX,footnote10}.            
                                                                
Here we will argue that in this new regime, at the initial
stage of the heavy ion collisions, a very dense 
non-equilibrium quark-gluon state will
be formed in the nucleus fragmentation region. Before addressing the
case of heavy ion collisions, we briefly discuss the case of the central 
$pA$ collisions. To visualize the space time picture of the BBL in the
proton-nucleus collisions, let us consider the  rest frame of  the nucleus.   
Within the semiclassical approximation valence partons keep the
momentum of energetic parent projectiles. This property of the infinite 
momentum frame  Schr$\rm \ddot{o}$dinger wave function 
is well known in the non-relativistic quantum mechanics,
in the pQCD approximations, in the parton model approach. 
It is valid also in BBL where a
strong 
PQCD
interaction is important for the
large intervals in rapidity only. But because of dominance in this case of 
the kind of two body kinematics such interaction does not change fraction
of projectile momentum carried by the valence partons. At the
same time the distribution of small $x$ bare particles is far from    
understood because the operator of the number of                             
bare particles is not the integral of motion in QCD. So, the spectrum of  
leading hadrons, the probability of the gap in rapidity between 
projectile, and target fragmentation may depend on atomic number. 
The semiclassical approximation agrees well with the conventional picture 
of high energy hadron-hadron collisions. Fast partons in  a nucleon with 
momentum $p_{N}$ are contracted to a pancake of transverse radius $r_{N}$ 
and a small longitudinal size $z=r_N m_N/p_N$ while the small $x$ partons 
form a pancake of $x_{V}/x$ times larger longitudinal  size 
($x_{V}\approx 0.2$ is the average $x$ of the valence quarks). The strong 
interaction of a nucleon with a target  occurs when the
longitudinal size of the fast nucleon becomes comparable with nucleon radius.
It is small $x$ partons of longitudinal size $\sim 1 fm$ which eventually 
interact with the target. Thus valence quarks and gluons of the 
projectile keep  practically the same  longitudinal momentum  during 
collision. It is the distribution of small x partons which is
influenced by the nucleus medium. This phenomenon is the essence of
the parton model, and of the 
Dokshitzer-Gribov-Lipatov-Altarelli-Parisi (DGLAP) 
 evolution of hard processes 
and current models for the energy losses \cite{Baier}. In the
next discussion, we will use this property of QCD.

A 
parton with sufficiently large $x$ belonging to a fast  proton emits a
virtual photon (hard gluon)  long before the target and it interacts
with the target, in a black regime releasing 
the fluctuation, e.g. a Drell-Yan pair.  This leads to a qualitative change 
in  the interactions for the partons with $x_p, p_t$ satisfying the          
condition that                                                              
\begin{equation}                                                  
x_A=4 p_t^2/(x_{p}s_{NN}),
\label{limit}
\end{equation}
is in the black body kinematics for the resolution scale 
$p_t\leq p_t^{b.b.l.}(x_A)$. Here, $ p_t^{b.b.l.}(x_A)$ is the 
maximum $p_t$ for 
which the black body approximation is applicable. In the kinematics of LHC, 
$Q^2\approx 4(p_t^{b.b.l.})^2$ can be estimated by using formulae derived in  
\cite{FKS96}.  At minimal $x_A$ $p_t^{b.b.l.}$ may reach 4 GeV/c.           
All the partons with such $x_p$ will obtain $p_t(jet)\sim  p_t^{b.b.l.}(x_A)$ 
leading to the multijet production. The blackbody regime will extend down in 
$x_p$ with increasing  the incident energy. For LHC, for $p_t \leq 3 GeV/c$,  
this regime may  cover the whole region of $x_p \geq 0.01$ where of the order 
of ten partons reside. Hence, in this limit most of the final states will   
correspond to multiparton collisions. For $p_t \leq 2 GeV/c$ the region 
extends to $x_p \geq 0.001$. At LHC 
fragmentation of such partons result in the
generation of hadrons at central rapidities. The dynamics of conversion of
the high $p_t$ partons with similar rapidities to hadrons is certainly
a collective effect which deserves a special consideration.

The total inclusive cross sections in $pA$ scattering can be calculated   
within BBL using a similar method to that used in $\gamma^*$ - nucleus
scattering \cite{bbl}. In particular, the total cross section of the 
dimuon production is:
\begin{equation}
{d\sigma(p+A \to \mu^+\mu^- + X)\over dx_Adx_p}=
{4\pi\alpha^2\over 9}{K(x_A,x_p,M^2)\over M^2}
F_{2p}(x_p,Q^2)\cdot {1\over 6\pi^2} M^2\cdot 2\pi R_A^2 \ln(x_0/x_A).
\label{DY}
\end{equation}
Here the $K$-factor has the same meaning as in the leading twist case, but 
$K-1$ should be smaller since it originates from the gluon emissions 
from the parton belonging to the proton only. $x_0$ is the maximal $x$
for which BBL is valid. For the smallest $x_A$ (forward kinematics) 
Eq.\ref{DY} may be valid  at LHC
for $ M^2 \leq 60 GeV^2$.
Obviously, Eq.\ref{DY}  prediction for  the $M^2$, $x_A$
dependence of the cross section are distinctively different 
from the DGLAP limit. This difference  
would be less pronounced in the case of $pp$ scattering where scattering 
at large impact parameters may mask the BBL contribution. 
Another signal for the onset of the BBL is 
a broadening of the $p_t$ distribution of the dimuons as compared to the 
DGLAP expectations, see \cite{Jamal} for a calculation of this effect
in  the color glass condensate model. This effect is similar to the case 
of $p_t$    
distribution of leading partons in the deep inelastic scattering \cite{bbl}.

As $x_A$ decreases further, the formulae for BBL will probably 
overestimate cross section because the interaction  with heavy 
nucleus of sea quarks and gluons in the projectile 
proton would become black as well.       
                                      
The onset of the BBL will lead also to gross changes in the hadron production: 
there is a much stronger drop with $x_F$ of the spectrum in the proton 
fragmentation region accompanied by a significant $p_t$ broadening of the 
spectrum. There is also  the enhancement of  hadron production, at smaller 
rapidities. Indeed, the individual partons in this limit are resolved    
at the virtuality scale corresponding to transverse momenta 
$\sim p_t^{b.b.l.}$ without losing  a finite fraction of the  
light cone momentum.  Hence, the limit of independent parton fragmentation  
\cite{berera} will be realized with an important amplification since 
the leading partons will have much larger transverse momenta \cite{DM} than 
that expected from the  the estimate of \cite{berera}  
based on the $p_t$ broadening observed at the fixed target energies.
We want to emphasize here that the the approximation of zero fractional losses 
holds in the  pQCD regime and appears to hold in the various models of the
onset of the BBL, see e.g. \cite{DM}. At RHIC this regime may hold for
the very forward hadrons and  could be checked \cite{neutrons}
by studying the production of leading hadrons 
in the central p($^2$H)A collisions \cite{footnote2}. 
The propagation of a proton interacting in the BBL  results in the removal of 
all partons in the nucleus with the proton impact parameter and    
$p_t \leq p_t^{b.b.l.}$. This leads to the formation of a 
$\sim 1 $ fm radius tube in a  perturbative phase.
Thus, the essence of the BBL is the striping of nucleons off the soft QCD  
modes and releasing the gas of free quarks and gluons with large
transverse momenta. Remember that the pQCD interactions  between
quarks and gluons within the same fragmentation region is small.
Thus, we conclude that the effects of the BBL should be first 
manifested in the projectile fragmentation region and should
gradually expand towards  central rapidities. The detectors with a 
forward acceptance would be optimal for this physics.

Let us now discuss the  nucleus fragmentation region 
in the central heavy ion  
collisions. It was discussed previously in the framework of the soft
dynamics, see \cite{Larry} and references therein, and a significant
 increase of the densities was found. Let us demonstrate now that even 
more striking effects are expected in the BBL regime.
In this case (in difference from $pA$ collisions) soft modes
will be stripped off in the whole volume of the nucleus.
 To calculate the properties 
of the quark - gluon state produced in the tube of radius $R_A$
in the target fragmentation 
region we first evaluate the main characteristics of the process in the 
rest frame of the fragmenting ion. The incoming nucleus is a pancake shape 
with the longitudinal length $\sim 1 fm$ for the soft interactions and
the longitudinal length $r_N (m_N/p_N)(x_V/x)$ for hard interactions.  
Hence, the nucleons at different locations along the  collision axis are hit 
one after another. In the BBL no spectators are left. The hit partons are 
produced with the same $x$ that they had in the nucleus (since the energy 
losses are $\propto 1/s$). We also have average, $p_t \sim p_t^{b.b.l.}$ 
and virtuality, $\mu^2\leq (p_t^{b.b.l.})^2$. The partons move in the 
direction of the projectile nucleus. Since they are emitted at finite 
angles their longitudinal velocity is smaller than the speed of light. 
Hence,  they are left behind the projectile wave. However, since the  
emission angles are  small the shock wave is formed compressing the 
produced system in the nucleus rest frame, see discussion below.

To estimate  the produced densities, we first calculate the emission angles.  
Since the parton's $x$ is not changed, 
\begin{equation}
(E_i-p_i^z )=xm_N,
\end{equation}
leading to 
\begin{equation}
p_z=(\mu^2+p_t^2)/2xm_N -xm_N/2 \approx p_t^2/2xm_N.
\end{equation}
Here in the last equation  we have neglected $\mu^2$  compared to  
$p_t^2$ which is legitimate in the leading order. Since $\mu^2\geq 0$,
neglected terms would increase $p_z$ making the emission angles, $\theta$,
even smaller. Thus in the BBL the angles $\theta$ 
\begin{equation}
\theta\simeq p_t/p_z\sim 2x m_N /p_t,
\end{equation}
are small. So  the length  of the  produced  
wave package is reduced from a naive value of 2 $R_A$ by a large factor 
\begin{equation}
S=1/(1-\cos\theta)\approx p_t^2 / 2x^2 m_N^2.
\label{suppr}
\end{equation}

However, we must also  take into account  that the products of the  
nucleon fragmentation as a whole move forward in the target rest
frame. The four-momentum of the system
can be calculated from the condition that this parton system carries
almost all of the  light-cone momentum of the initial nucleon:   
\begin{equation}
(\sqrt{M^2+p_z^2}-p_z)/m_N=1,
\end{equation}
where $M^2=\sum_i p_{i,~t}^2/x_i$ is the square of the invariant mass
of the system. Hence, $p_z=M^2/2m_N$, and the Lorentz factor 
\begin{equation}
\gamma=E/M=\sqrt{M^2+(M^2/2m_N)^2}/M \approx M/2m_N.
\label{gamma}
\end{equation}
Combining Eqs.\ref{suppr}, \ref{gamma}
we find for the decrease of volume relative to the  nucleus volume: 
 \begin{equation}  
D=(2m_N/M)\cdot  <p_t^2/ 2m_N^2 x^2>.
\label{9}
\end{equation}
First, to make simple numerical estimates we assume  that all relevant 
partons carry equal light cone fractions 1/N. (The 
dispersion in $x_i$ leads for a further decrease of the volume
and hence to a  larger density  of the produced
system, see below.) In  this case we obtain
\begin{equation}
D=M/m_N=Np_t/m_N.
\end{equation}
Thus,  we have demonstrated that the volume is indeed significantly 
smaller than the nuclear volume.
 
The next step is to estimate the energy per unit volume, $R_E$. It is 
convenient to present it in units the energy density of  nuclear matter 
which equals 0.16 GeV/fm$^3$.  The lower limit on $R_E$ is obtained by
neglecting the dispersion in $x_i$ and taking $x_i=1/N$,  
\begin{equation}   
R_E =D \cdot  Np_t/m_N = N^2p_t^2/m_N^2.
\end{equation}
At LHC for $x\geq 0.01$ the BBL extends up to $p_t=2 \div 3$ GeV/c,    
and the number of partons for such x and virtualities is  $\sim 8 \div 10$.
This leads to 
\begin{equation}
R_E > 250. 
\end{equation}
If we take into account the dispersion in $x_i$ we find:
\begin{equation}
R_E = \sum_i p_{i~t}^2/m_N^2 x_i^2.
\label{13}
\end{equation}
Though in the first approximation the dispersion can be neglected for       
valence quarks and probably for the gluons separately, the quarks
carry an average $x$ that is 
a factor of two  larger than gluons.
This leads to  a further increase in 
the energy density. Taking $N_q=3, x_q=1/6, N_g=6, x_g=1/12$, 
we find a rather modest increase of $R_E$ (a factor of 4/3) for the 
same $N=9$, corresponding to  the energy densities $\geq 50 GeV/fm^3$.
(If we assume proximity of BBL  at  RHIC for
the  fragmentation region for   $p_t\sim 1 GeV/c$,
we find quark-gluon energy densities $\sim 10 GeV/fm^3$.)\cite{comment}
More significantly, the difference in
average x's of quarks and gluons leads to a different direction of the
flow of the quarks and gluons in the c.m. frame of the produced
system. For the above numerical example,  $k_3/k_t \sim 0.6 $ for    
quarks and  $\sim -0.3$ for gluons. Obviously, this pattern       
will enhance the interactions of quarks and gluons at the next stage   
of the interactions, making equilibration more likely.

Let us briefly discuss these interactions and the
possible experimental signals. 
It follows from Eqs.\ref{9} - \ref{13}
that at LHC in the first stage of
collisions a strongly compressed hot  quark-gluon state of the 
ellipsoidal shape is formed with the small principal
axis of $\sim 1 fm$ and density  $\rho \geq 25$ partons per $fm^3$.
At the higher rapidity end, this ellipsoid boarders essentially parton free
space; on the end close to central rapidities, it boarders a hot 
$q\bar q g$ state.
The scattering length for parton $i$ can be estimated as $l_i=1/(\sum_j
N_j \sigma_{ij})$, corresponding to the scattering length being smaller
than 1 fm for $\sigma\ge 0.5 mb$. To estimate the interaction cross
section, 
we note that the average invariant energy $s\approx 2p_t^2\sim 8
GeV^2$. The  initial stage of reinteractions  certainly is a highly
non-equilibrium process. Nevertheless to do a perturbative estimate 
 we can conservatively introduce a cutoff
on the momentum transfer $p\sim {\pi\over 2}\rho^{-1/3}$, leading to the
leading order estimate for the gluon - gluon cross section $\geq 1
mb$. Nonperturbative effects, which remain strong in the gluon sector
up to $\sqrt s \sim 3$ GeV, are likely to increase these interactions further. 
Consequently, we expect partons to rescatter strongly at the second
stage, though much more detailed modeling is required 
to find out whether the system may reach thermal equilibrium.
The large angle rescatterings of partons
will lead to production of partons at higher rapidities
and re-population of the cool region. In particular, 
two gluons have the right energies to produce near threshold $c\bar c$
pairs and in particular $\chi_c$-mesons with rather
small transverse momenta and $x_F(c\bar c)\sim 2x_g \sim 0.1$. 
Also leading photons can
be produced in the $q g \to \gamma q$ subprocesses, though in
difference from the central region the $q\bar q \to \mu^+\mu^-$
production will be
suppressed due to the lack of antiquarks.  Another high density effect
is production of leading 
nucleons via recombination of quarks 
with subsequent escape to the cool region. Hence we expect a rather
paradoxical situation that the production of leading hadrons in $AA$
collisions will be stronger than in the central pA collisions.

To summarize, we have demonstrated that the onset of the black body    
regime in the interactions in the target fragmentation region which is
likely at LHC for 
a large range of virtualities,  will lead to the formation of   a
new superdense initial state in the nuclei fragmentation region with densities
exceeding nuclear densities at least by a factor of 300.
Our reasoning is however insufficient to demonstrate whether
thermalization processes will be strong enough for the system to reach
equilibrium necessary for  formation of
metastable states - new QCD phases suggested in a number of recent papers, 
see the review in \cite{Alford:2001dt}.

Further studies of the experimental manifestations of the formation of
high density quark-gluon states are needed as well as an investigation
of hard processes in the central proton-nucleus collisions which will allow a
determination  of whether  this new state of valence quark-gluon matter  
can be formed  at RHIC energies. Similar effects, like correlation of 
high $p_t$ hadron production in the fragmentation regions, should be 
present in the central $pp$ collisions at LHC
(where jet production at central rapidities is used as a trigger of the
centrality
\cite {FELIX}).

We thank D.Kharzeev, A.Mueller, E.Shuryak, D.Son, R.Venugopalan
for useful discussions and  GIF and DOE for support.


\begin{thebibliography}{99}
\bibitem{footnote1} The assumption that all hadron wave
packages at sufficiently high energies interact with  heavy nuclei with 
maximal possible cross section of $2\pi R_A^2$ was made in the preQCD 
studies in \cite{Gribov}. This assumption was later  questioned
\cite{Bj} because it contradicts to Bjorken scaling  which is a
consequence of the color transparency phenomenon in LO pQCD
\cite{FS}. However, in  the small x limit both  the color transparency
phenomenon for the spatially small quark gluon configurations of fixed
transverse size as well as the Bjorken scaling disappear.
\bibitem{footnote10} Physics of forward collisions at RHIC is not far 
from BBL for quarks and gluons with
$x \geq 0.1$ for $p_t \leq 1 \div 2 GeV$.
But the theoretical uncertainties in this case are significantly larger.
\bibitem{comment} The densities we find are at least a factor of 10
higher than in \cite{Larry} due to much larger release of energy in
the BBL and due to significantly  larger  longitudinal compression
of the interaction volume.
Our estimate neglects conversion of the released partons into hadrons 
before they reach the back edge of the fragmenting nucleus. 
If $p_t^{b.b.l.}$ is small enough, this effect may become important.
\end{thebibliography}
\end{document}